\documentclass[12pt]{iopart}

\usepackage{graphicx}
\usepackage{color}
\usepackage{mathrsfs}
\usepackage{bm}
\usepackage{bbm}
\usepackage{amsfonts}
\begin{document}

\title{Non-Markovian entanglement dynamics between two coupled qubits in the same environment}
\author{Wei Cui$^{1,2}$, Zairong Xi$^{1 *}$ and Yu Pan$^{1,2}$}
\address{%
$^1$Key Laboratory of Systems and Control, Institute of Systems
Science, Academy of Mathematics and Systems Science, Chinese Academy
of Sciences, Beijing 100190, People's Republic of China

$^2$Graduate University of Chinese Academy of Sciences, Beijing
100039, People's Republic of China
}%
\ead{zrxi@iss.ac.cn}

\begin{abstract}
We analyze the dynamics of the entanglement in two independent
non-Markovian channels. In particular, we focus on the entanglement
dynamics as a function of the initial states and the channel
parameters like the temperature and the ratio $r$ between $\omega_0$
the characteristic frequency of the quantum system of interest, and
$\omega_c$ the cut-off frequency of Ohmic reservoir.  We give a
stationary analysis of the concurrence and find that the dynamic of
non-markovian entanglement concurrence $\mathcal{C}_{\rho}(t)$ at
temperature $k_BT=0$ is different from the $k_BT>0$ case.  We find
that ``entanglement sudden death" (ESD) depends on the initial state
when $k_BT=0$, otherwise the concurrence always disappear at finite
time when $k_BT>0$, which means that ESD must happen. The main
result of this paper is that the non-Markovian entanglement dynamic
is fundamentally different from the Markovian one. In the Markovian
channel, entanglement decays exponentially and vanishes only
asymptotically, but in the non-Markovian channel the concurrence
$\mathcal{C}_{\rho}(t)$ oscillates, especially in the high
temperature case. Then an open-loop controller adjusted by the
temperature is proposed to control the entanglement and prolong the
ESD time.
\end{abstract}
\pacs{03.65.Ud, 03.65.Yz, 03.67.Mn, 05.40.Ca}
\maketitle

\section{Introduction}
Entanglement is  a remarkable feature of quantum mechanics, and its
investigation is both of practical and theoretical significance. It
is viewed as a basic resource for quantum information processing
(QIP) \cite{Nielsen}, like realizing high-speed quantum computation
\cite{Bennett} and high-security quantum communication
\cite{Bouwmeester}. It is also a basic issue in understanding the
nature of nonlocality in quantum mechanics \cite{Einstein, Bell,
Bennett2}. However, a quantum system used in quantum information
processing inevitably interacts with the surrounding environmental
system (or the thermal reservoir), which induces the quantum world
into classical world \cite{H.P.Breuer,Zurek,zurek}. Thus, it is an
important subject to analyze the entanglement decay induced by the
unavoidable interaction  with the environment
\cite{weiss,MY,Cui2,Zhang09,HPC}. In one-party quantum system, this
process is called decoherence
\cite{Everett,Zurek,GG,MG,HH,Cui,Zhang1}. In this paper, we will
analyze the entanglement dynamics of bipartite non-Markovian quantum
system. As well known, the system can only couple to a few
environmental degrees of freedom for short times. These will act as
memory. In short time scales environmental memory effects always
appear in experiments \cite{Rodriguez}. The characteristic time
scales become comparable with the reservoir correlation time in
various cases, especially in high-speed communication. Then an
exactly analytic description of the open quantum system dynamic is
needed, such as quantum Brownian motion(QBM) \cite{zurek}, a
two-level atom interacting with a thermal reservoir with Lorentzian
spectral density \cite{Garraway}, and the devices based on solid
state \cite{Chirolli} where memory effects are typically
non-negligible. Due to its fundamental importance in quantum
information processing and quantum computation, non-Markovian
quantum dissipative systems have attracted much attention in recent
years \cite{H.P.Breuer, Hu,Prager,Lorenz,Cao, Bellomo}. Recently,
researches on quantum coherence and entanglement influenced and
degraded by the external environment become more and more popular,
most of the works contributed to extend the open quantum theory
beyond the Markovian approximation \cite{Liu, SM, Maniscalco2}. In
\cite{Liu}, two harmonic oscillators in the quantum domain were
studied and their entanglement evolution investigated with the
influence of thermal environments. In \cite{SM}, the dynamics of
bipartite Gaussian states in a non-Markovian noisy channel were
analyzed. All in all, non-Markovian features of system-reservoirs
interaction have made great progress, but the theory is far from
completion, especially how the non-Markovian environmental influence
the system and what the difference is between Markovian and
non-Markovian system evolution are not clear.

In this paper we will compare the non-Markovian entanglement
dynamics with the Markovian one \cite{Zhang2} in Ohmic reservoir
with Lorentz-Drude regularization in the following three conditions:
$\omega_0\ll\omega_c$, $\omega_0\approx\omega_c$ and
$\omega_0\gg\omega_c$, where $\omega_0$ is the characteristic
frequency of the quantum system of interest and $\omega_c$ the
cut-off frequency of Ohmic reservoir. Thus, $\omega_c\ll\omega_0$
implies that the spectrum of the reservoir does not completely
overlap with the frequency of the system oscillator and
$\omega_0\gg\omega_c$ implies the converse case. Another point of
the entanglement dynamics is the temperature. We characterize our
system by low temperature, $k_BT=0.03\omega_0$, medium temperature,
$k_BT=3\omega_0$, and the high temperature $k_BT=300\omega_0$. We
give stationary analysis of the concurrence \cite{MY} and find that
the dynamics of
 non-markovian
 entanglement concurrence $\mathcal{C}$ at temperature $k_BT=0$  is fundamentally different from the
 $k_BT>0$. We find that ``entanglement sudden death" (ESD) depends on the initial state
when $k_BT=0$, otherwise the concurrence always disappear at finite
time when $k_BT>0$, which means that ESD must happen.
 Maniscalco S et.al studied
the separability function $S(\tau)$ in \cite{SM}, where the
entanglement oscillation appears for twin-beam state in
non-Markovian channels for high temperature reservoirs. The main
result of this paper is that the non-Markovian entanglement dynamics
is fundamentally different from the Markovian one. In the Markovian
channel, entanglement decays exponentially and vanishes only
asymptotically, but in the non-Markovian channel the concurrence
$C_{\rho}(t)$ oscillates, especially in the high temperature case.

The paper is organized as follows. We first introduce the open
quantum system and the non-Markovian quantum master equation for
driven open quantum systems by the noise and dissipation kernels. In
Sec. III we introduce the Wootters' concurrence and the initial ``X"
states. By substituting the initial states into the master equation
we get the first order coupled differential equations, and give the
stationary analysis. In Sec. IV, we numerically analyze the
Markovian and non-Markovian entanglement dynamics. Then an open-loop
controller adjusted by the temperature is proposed to control the
entanglement and prolong the ESD time. Conclusions and prospective
views are given in Sec. V.
\section{The model}
Our system consists of a pair of two-level atoms (two qubits)
equally and resonantly, coupled to a single cavity mode, with the
same coupling strength. The master equation for the reduced density
matrix $\rho(t)$ which describes its dynamics is given by
\cite{H.P.Breuer, Cui, SM,Maniscalco2,Intravaia}
\begin{eqnarray}\fl
\frac{d\rho(t)}{dt}=\frac{\Delta(t)+\gamma(t)}{2}\sum_{j=1}^2\{2\sigma_{j}^-
\rho\sigma_{j}^{+}-\sigma_{j}^{+}\sigma_{j}^-\rho-\rho\sigma_{j}^{+}\sigma_{j}^-\}\nonumber\\+
\frac{\Delta(t)-\gamma(t)}{2}\sum_{j=1}^2\{2\sigma_{j}^{+}\rho\sigma_{j}^--\sigma_{j}^-\sigma_{j}^{+}\rho
-\rho\sigma_{j}^-\sigma_{j}^{+}\}.
\end{eqnarray}
where
$\sigma^+=\frac{1}{2}(\sigma_1+i\sigma_2),~~\sigma^-=\frac{1}{2}(\sigma_1-i\sigma_2)$,
with $\sigma_1$, $\sigma_2$ the Pauli matrices. The time dependent
coefficients appearing in the master equation can be written, to the
second order in the coupling strength, as follows
\begin{equation}
\label{DDC}
 \begin{array}{rcl}
\Delta(t)&=&\int_0^td\tau k(\tau)\cos(\omega_0\tau),\\
\gamma(t)&=&\int_0^td\tau \mu(\tau)\sin(\omega_0\tau),
 \end{array}
 \end{equation}
with
\begin{equation}
 \begin{array}{rcl}
k(\tau)&=&2\int_0^{\infty}d\omega
J(\omega)\coth[\omega/2k_BT]\cos(\omega \tau),\\
\mu(\tau)&=&2\int_0^{\infty}d\omega J(\omega)\sin(\omega \tau),
 \end{array}
 \end{equation}
being the noise and the dissipation kernels, respectively. This
master equation (1) is valid for arbitrary temperature. The
coefficient $\gamma(t)$ gives rise to a time dependent damping term,
while $\Delta(t)$ the diffusive term. The non-Markovian character is
contained in the time-dependent coefficients, which contain all the
information about the short time system-reservoir correlations
\cite{H.P.Breuer}. In the previous equations $J(\omega)$ is the
spectral density characterizing the bath,
\begin{equation}
J(\omega)=\frac{\pi}{2}\sum_i\frac{k_i}{m_i\omega_i}\delta(\omega-\omega_i)
 \end{equation}
 and the index $i$ labels the different field mode of
the reservoir with frequency $\omega_i$. Let the Ohmic spectral
density with a
 Lorentz-Drude cutoff function,
\begin{equation}
J(\omega)=\frac{2}{\pi}\omega\frac{\omega_c^2}{\omega_c^2+\omega^2},
 \end{equation}
 where $\omega$ is the frequency of the bath, and $\omega_c$ is the high-frequency
 cutoff.

 Then the closed analytic expressions for $\Delta(t)$ and
 $\gamma(t)$ are \cite{Cui, Maniscalco2}
  \begin{equation}
 \gamma(t)=\frac{\omega_0r^2}{1+r^2}[1-e^{-r\omega_0t}\cos(\omega_0t)-re^{-r\omega_0t}\sin(\omega_0t)],
 \end{equation}

\begin{eqnarray}
\fl \Delta(t)=\omega_0\frac{r^2}{1+r^2}\{\coth(\pi r_0)-\cot(\pi
r_c)e^{-\omega_ct}[r\cos(\omega_0t)-\sin(\omega_0t)]
\nonumber\\
+\frac{1}{\pi
r_0}\cos(\omega_0t)[\bar{F}(-r_c,t)+\bar{F}(r_c,t)-\bar{F}(ir_0,t)-\bar{F}(-ir_0,t)]\nonumber\\
-\frac{1}{\pi}
\sin(\omega_0t)[\frac{e^{-\nu_1t}}{2r_0(1+r_0^2)}[(r_0-i)\bar{G}(-r_0,t)+(r_0+i)\bar{G}(r_0,t)]\nonumber\\
+\frac{1}{2r_c}[\bar{F}(-r_c,t)-\bar{F}(r_c,t)]]\},
\end{eqnarray}
where $r_0=\omega_0/2\pi k_BT$, $r_c=\omega_c/2\pi k_BT$,
$r=\omega_c/\omega_0$, and
 \begin{equation}
\bar{F}(x,t)\equiv _2F_1(x,1,1+x,e^{-\nu_1t}),
 \end{equation}
  \begin{equation}
\bar{G}(x,t)\equiv _2F_1(2,1+x,2+x,e^{-\nu_1t}).
 \end{equation}
$\nu_1=2\pi k_BT$, and $ _2F_1(a,b,c,z)$ is the hypergeometric
function.   Note that, for time $t$ large enough, the coefficients
$\Delta(t)$ and $\gamma(t)$ can be approximated by their Markovian
stationary values $\Delta_M=\Delta(t\rightarrow\infty)$ and
$\gamma_M=\gamma(t\rightarrow\infty)$. From Eqs. (6) and (7) we have
\begin{equation}
\gamma_M=\frac{\omega_0r^2}{1+r^2},
 \end{equation}
 and \begin{equation}
\Delta_M=\omega_0\frac{r^2}{1+r^2}\coth(\pi r_0).
 \end{equation}
 \begin{figure*}

\centerline{\scalebox{1}[1]{\includegraphics{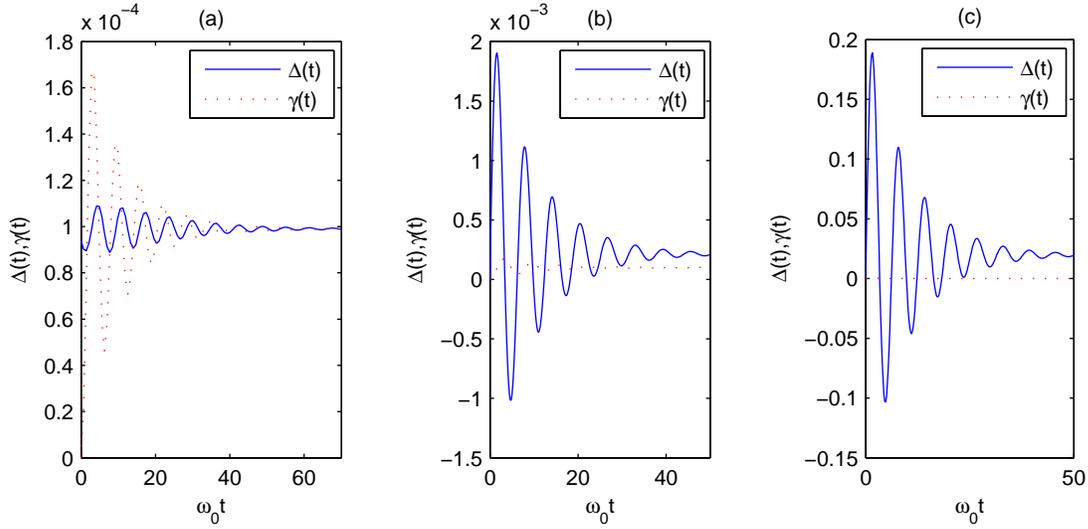}}}
\caption{(Color online) Dynamics of non-Markovian coefficients
$\Delta(t)$ (blue solid line) and $\gamma(t)$ (red dotted line) at
different temperatures: (a)low temperature $k_BT=0.01$, (b)medium
temperature $k_BT=1$, and (c)high temperature $k_B(t)=100$,
respectively. The other coefficients are chosen as $r=0.1$
$\omega_0=1$, and $\alpha^2=0.01$. }
\end{figure*}
Note that $\gamma(t)$ has nothing to do with the temperature
\cite{Intravaia}. In Fig.1 we plot the time evolution of
non-Markovian coefficients $\Delta(t)$ and $\gamma(t)$ in different
channel temperatures. In Fig. 1(a), the temperature is $k_BT=0.01$.
There are two important main points embodied in the Figure, the
first is that the coefficient $\gamma(t)$ has dominated the system
dissipation at low temperature, the other $\Delta_M=\gamma_M$ in the
long time limit. Fig. 1(b) and (c) are the evolution at the medium
temperature and high temperature respectively. The Figure shows that
the larger the temperature, the more important the coefficient
$\Delta(t)$.

\section{Concurrence and initial states}

In order to describe the entanglement dynamics of the bipartite
system, we use the Wootters concurrence \cite{MY,Wootters}. For a
system described by the density matrix $\rho$, the concurrence
$\mathcal{C}(\rho)$ is
\begin{equation}
\mathcal{C}(\rho)=\max(0,\sqrt{\lambda_1}-\sqrt{\lambda_2}-\sqrt{\lambda_3}-\sqrt{\lambda_4}),
\end{equation}
 where $\lambda_1, \lambda_2, \lambda_3$, and $\lambda_4$ are the
 eigenvalues (with $\lambda_1$ the largest one) of the ``spin-flipped"
 density operator $\zeta$, which is defined by
\begin{equation}
\zeta=\rho(\sigma_y^A\otimes\sigma_y^B)\rho^{*}(\sigma_y^A\otimes\sigma_y^B),
\end{equation}
 where $\rho^{*}$ denotes the complex conjugate of $\rho$ and
 $\sigma_y$ is the Pauli matrix. $\mathcal{C}$ ranges in
 magnitude from 0 for a disentanglement state to 1 for a maximally entanglement state. The concurrence is related to the
 entanglement of formation $E_f(\rho)$ by the following relation \cite{Wootters}
\begin{equation}
E_f(\rho)=\varepsilon[\mathcal{C}(\rho)],
 \end{equation}
 where
\begin{equation}
\varepsilon[\mathcal{C}(\rho)]=h[\frac{1+\sqrt{1-\mathcal{C}^2(\rho)}}{2}],
 \end{equation}
and
\begin{equation}
h(x)=-x\log_2x-(1-x)\log_2(1-x).
 \end{equation}

 Assume that the system is initially an ``X" state, which has non-zero elements
 only along the main diagonal and anti-diagonal. The general
 structure of an ``X" density matrix is as follows
\begin{equation}
\hat{\rho}=\left(\begin{array}{cccc}
\rho_{11}&0&0&\rho_{14}\\
0&\rho_{22}&\rho_{23}&0\\
0&\rho_{23}^*&\rho_{33}&0\\
\rho_{14}^*&0&0&\rho_{44}
\end{array}\right).
 \end{equation}
 Such states are general enough to include states such as the Werner
 states, the maximally entangled mixed states (MEMSs) and the
 Bell states; and it also arises in a wide variety of physical
 situations \cite{Hagley, Bose, Pratt}. This particular form of the density matrix allows us to
 analytically express the concurrence as \cite{Yu4}
 \begin{equation}
 \mathcal{C}_{\hat{\rho}}^{X}=2\max\{0,K_1,K_2\},
 \end{equation}
 where
 \begin{equation}
  \begin{array}{rcl}
 K_1&=&|\rho_{23}|-\sqrt{\rho_{11}\rho_{44}},\\
 K_2&=&|\rho_{14}|-\sqrt{\rho_{22}\rho_{33}}.
 \end{array}
 \end{equation}
A remarkable aspect of the ``X" states is that the time evolution of
the master equation (1) is maintained during the evolution.
Substituting (17) into (1),
 the non-markovian master equation of two-qubits system, we obtain the following first-order coupled differential
 equations:
\begin{equation}\fl
 \begin{array}{rcl}
\dot{\rho}_{11}(t)&=&-2(\Delta(t)+\gamma(t))\rho_{11}(t)+(\Delta(t)-\gamma(t))\rho_{22}(t)+(\Delta(t)-\gamma(t))\rho_{33}(t),\\
\dot{\rho}_{22}(t)&=&(\Delta(t)+\gamma(t))\rho_{11}(t)-2\Delta(t)\rho_{22}(t)+(\Delta(t)-\gamma(t))\rho_{44}(t),\\
\dot{\rho}_{33}(t)&=&(\Delta(t)+\gamma(t))\rho_{11}(t)-2\Delta(t)\rho_{33}(t)+(\Delta(t)-\gamma(t))\rho_{44}(t),\\
\dot{\rho}_{44}(t)&=&(\Delta(t)+\gamma(t))\rho_{22}(t)+(\Delta(t)+\gamma(t))\rho_{33}(t)-2(\Delta(t)-\gamma(t))\rho_{44}(t),\\
\dot{\rho}_{23}(t)&=&-2\Delta(t)\rho_{23}(t),\\
\dot{\rho}_{14}(t)&=&-2\Delta(t)\rho_{14}(t).
\end{array}
 \end{equation}

From Eq. (18) the concurrence $\mathcal{C}$ is dependent on the
coefficients $\Delta(t\rightarrow\infty)$ and
$\gamma(t\rightarrow\infty)$ in the asymptotic long time limit. Eqs.
(10) and (11) give the stationary value of $\gamma(t)$ and
$\Delta(t)$, the Markovian limit
$$
\gamma_M\equiv\gamma(t\rightarrow\infty)=\frac{\omega_0r^2}{1+r^2},
$$
 and $$
\Delta_M\equiv\Delta(t\rightarrow\infty)=\omega_0\frac{r^2}{1+r^2}\coth(\frac{\omega_0}{2k_BT}).
 $$
 $\gamma_M$ doesn't depend on temperature, but $\Delta_M$ is monotonically increasing with respect to temperature
 $T$. When $T\rightarrow0$, $\Delta_M\rightarrow
 \frac{\omega_0r^2}{1+r^2}$. Noting $\coth(\pi
r_0)\simeq1+\frac{1}{\pi r_0}
 \simeq\frac{2k_BT}{\omega_0},$ at high temperature
  \begin{equation}
\Delta_M^{HT}=2k_BT\frac{r^2}{1+r^2}.\end{equation} So
$\Delta_M>\gamma_M$ is noticeable when temperature $k_BT>0$. From
Eqs. (20) we can get the stationary solution
 \begin{equation}
   \begin{array}{rcl}
\rho_{11}(t\rightarrow\infty)&=&\frac{\Delta_M-\gamma_M}{\Delta_M+\gamma_M}\rho_{33}(t\rightarrow\infty),\\
\rho_{22}(t\rightarrow\infty)&=&\rho_{33}(t\rightarrow\infty),\\
\rho_{33}(t\rightarrow\infty)&=&\frac{\Delta_M^2-\gamma_M^2}{4\Delta_M^2},\\
\rho_{44}(t\rightarrow\infty)&=&\frac{\Delta_M+\gamma_M}{\Delta_M-\gamma_M}\rho_{33}(t\rightarrow\infty).
   \end{array}
 \end{equation}
and
 \begin{equation}
   \begin{array}{rcl}
\rho_{23}(t\rightarrow\infty)&=&0,\\
\rho_{14}(t\rightarrow\infty)&=&0.
   \end{array}
 \end{equation}
 According to Eqs. (18, 19),
 \begin{equation}
 K_{1,2}(t\rightarrow\infty)=0-\frac{\Delta_M^2-\gamma_M^2}{4\Delta_M^2}<0.
 \end{equation}
This means that entanglement must disappear in a finite time period,
i.e. the ESD must happen.

When temperature $k_BT=0$, $\Delta_M\approx\gamma_M$. From Eqs. (20)
we can also get the stationary solution
 \begin{equation}
   \begin{array}{rcl}
\rho_{11}(t\rightarrow\infty)&=&0,\\
\rho_{22}(t\rightarrow\infty)&=&0,\\
\rho_{33}(t\rightarrow\infty)&=&0,\\
\rho_{44}(t\rightarrow\infty)&=&1.
   \end{array}
 \end{equation}
and
 \begin{equation}
   \begin{array}{rcl}
\rho_{23}(t\rightarrow\infty)&=&0,\\
\rho_{14}(t\rightarrow\infty)&=&0.
   \end{array}
 \end{equation}
 From Eqs. (18, 19),
 \begin{equation}
 K_{1,2}(t\rightarrow\infty)=0.
 \end{equation}
This means that entanglement maybe disappear asymptotically, or
oscillates, or other complex behaviors. In the following, we use the
numerical methods to demonstrate the concurrence evolution for a
special kind of ``X" state, the $\rho_{YE}$ state.

\section{Non-Markovian vs. Markovian entanglement dynamics}
\begin{figure*}
\centerline{\scalebox{0.8}[0.5]{\includegraphics{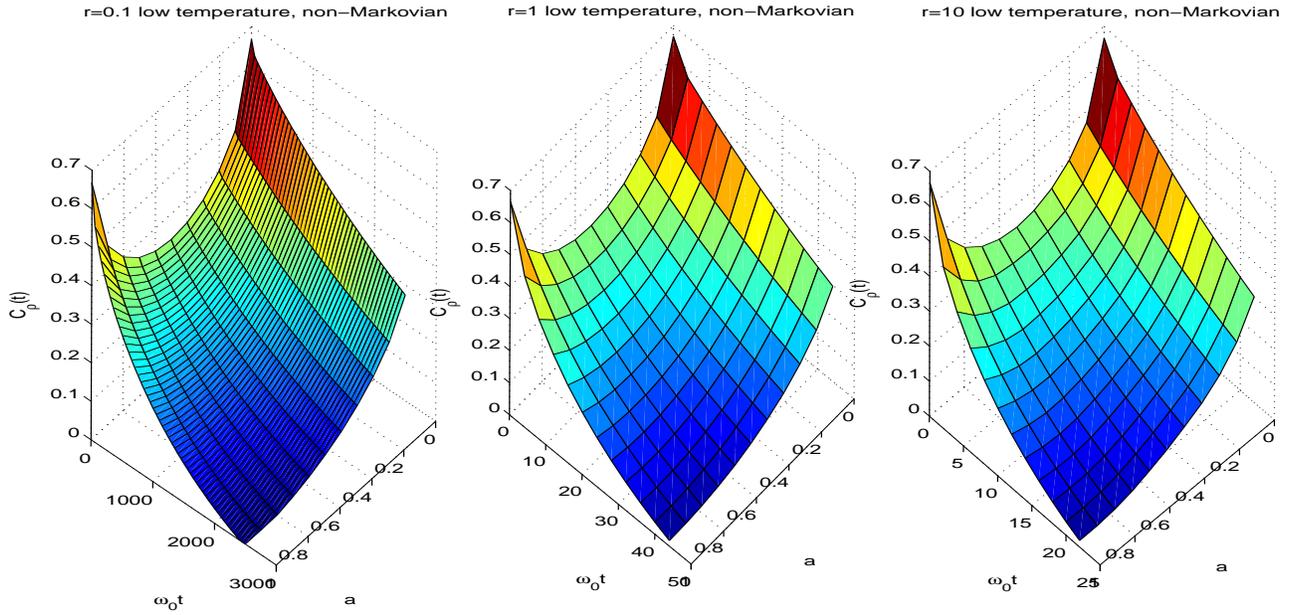}}}
\caption{(Color online)Time evolution of non-Markovian concurrence
as a function of parameter ``$a$"  in the low temperature
reservoirs.}
\end{figure*}

\begin{figure*}
\centerline{\scalebox{0.8}[0.5]{\includegraphics{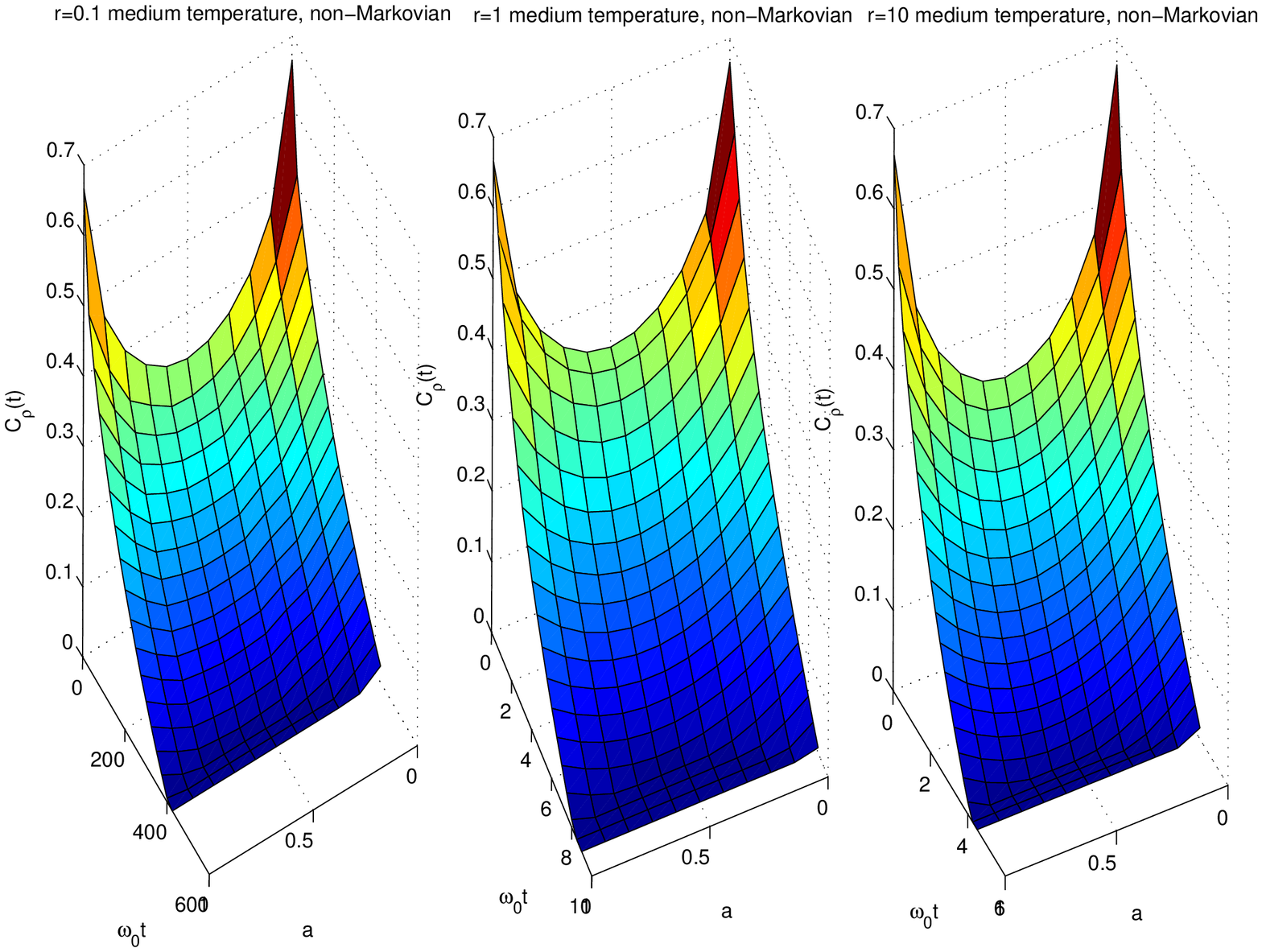}}}
\caption{(Color online)Time evolution of non-Markovian concurrence
as a function of parameter ``$a$"  in the medium temperature
reservoirs.}
\end{figure*}

\begin{figure*}
\centerline{\scalebox{0.5}[0.5]{\includegraphics{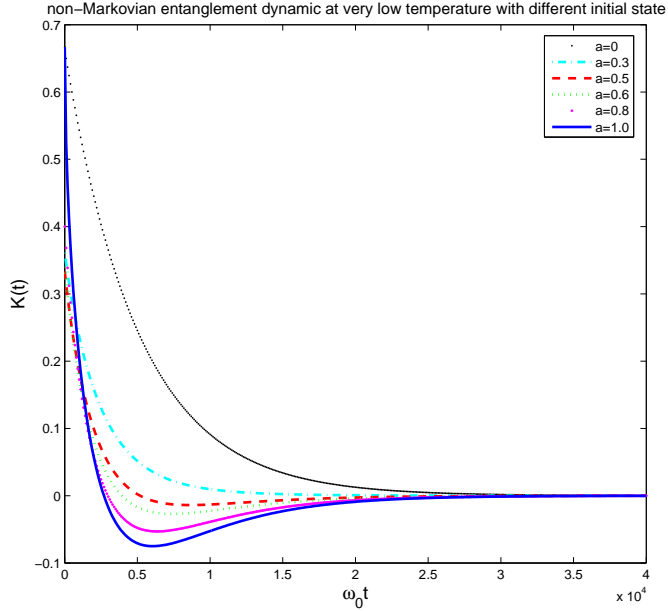}}}
\caption{(Color online)Time evolution of $K(t)$ for temperature
$k_BT=0.000001\omega_0$, $r=0.1$, and initial state
$\hat{\rho}_{YE}$ for the cases $a=0$ (black dotted), $a=0.3$ (cyan
dash-dotted line), $a=0.5$ (red dash line), $a=0.6$ (green
dotted-dotted line), $a=0.8$ (magenta asterisk), and $a=1.0$ (blue
solid line).}
\end{figure*}

\begin{figure*}
\centerline{\scalebox{0.8}[0.7]{\includegraphics{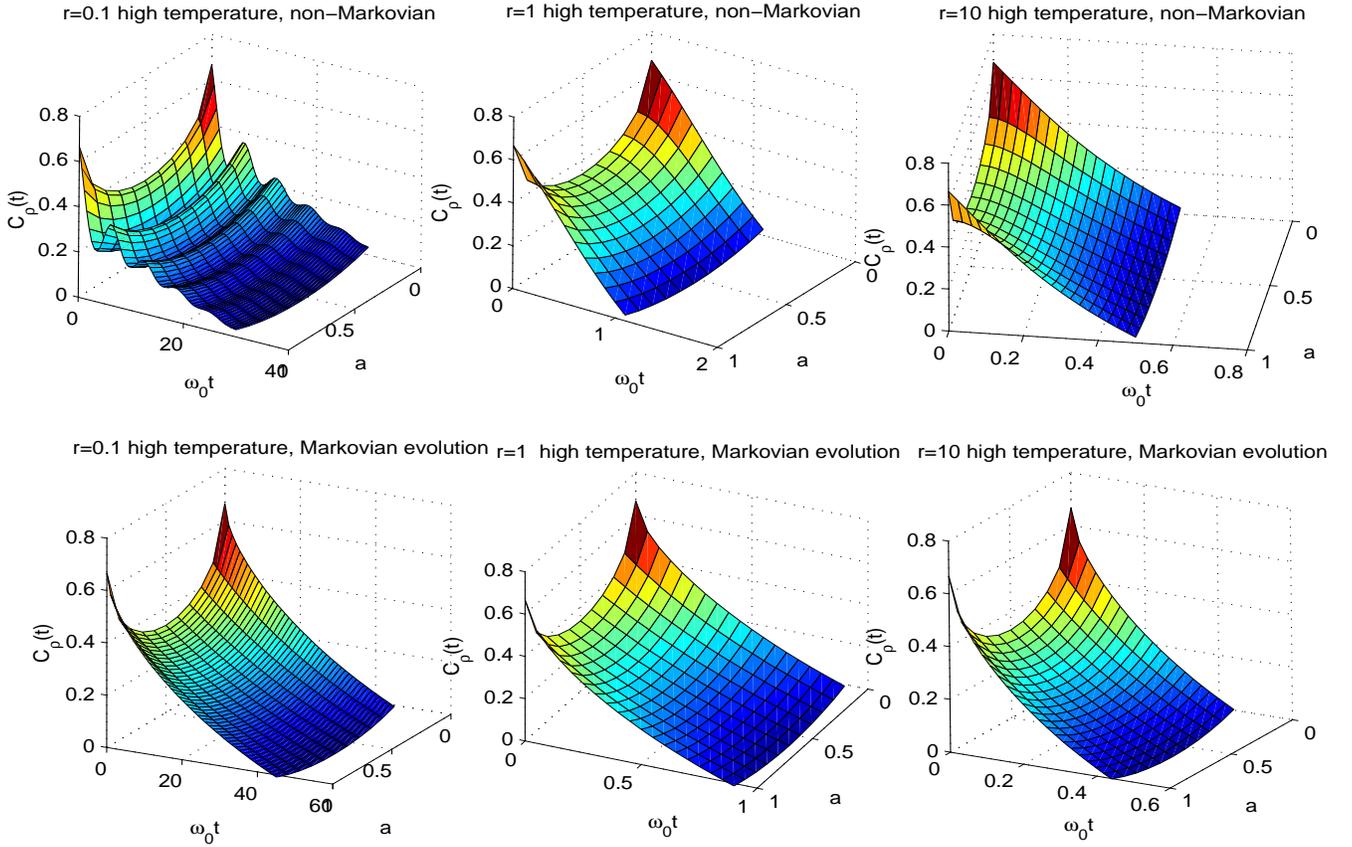}}}
\caption{(Color online)Comparing the non-Markovian entanglement
dynamics with the Markovian one by the time evolution of concurrence
as a function of parameter ``a" in high temperature reservoirs, at
$r=0.1$, $r=1$, $r=10$ respectively.}
\end{figure*}

\begin{figure*}
\centerline{\scalebox{0.8}[0.5]{\includegraphics{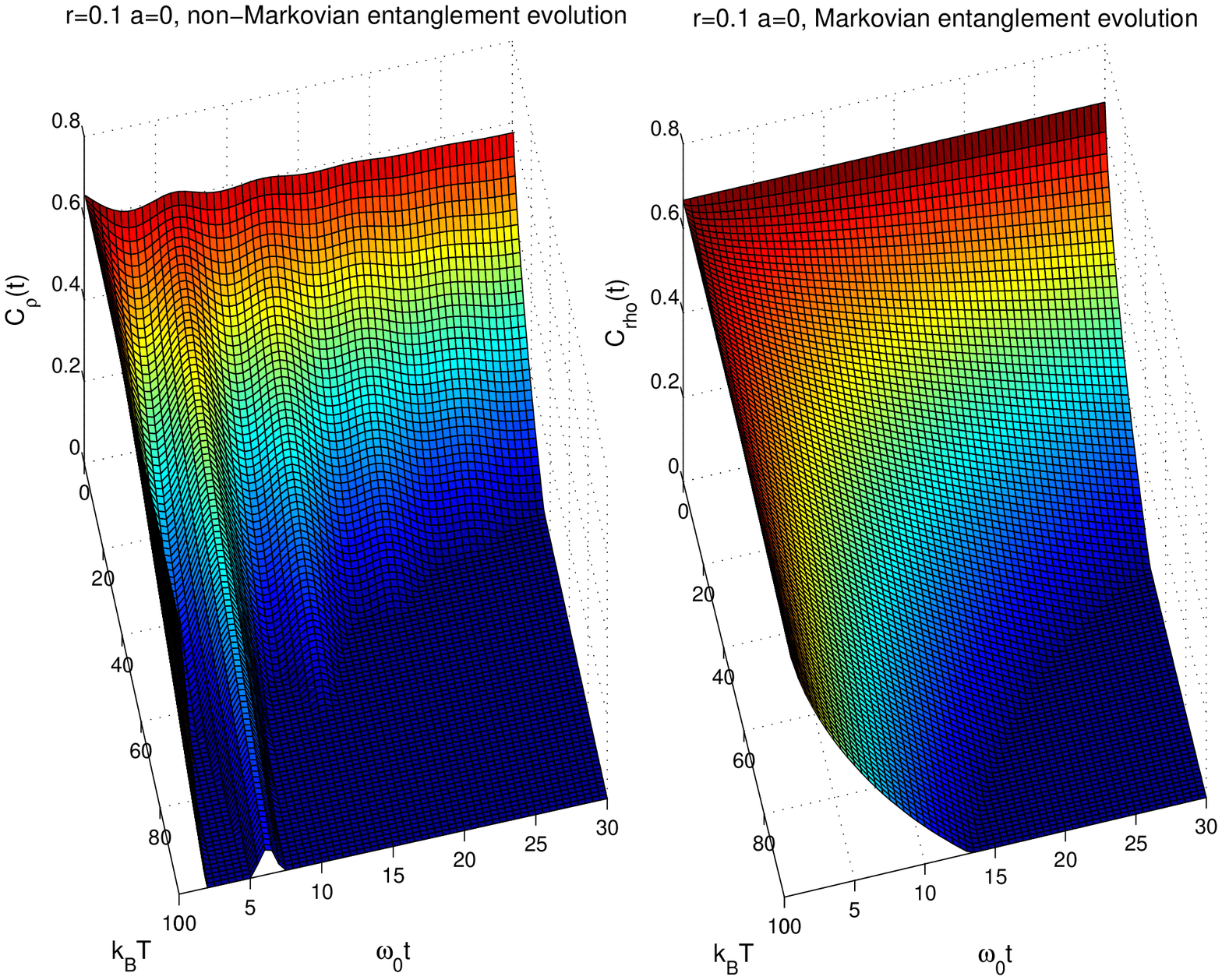}}}
\caption{(Color online)Comparing the non-Markovian entanglement
dynamics with the Markovian one by the time evolution of
$\mathcal{C}_{\rho}(t)$  as a function of  ``$k_BT$" for initial
state $a=0$ and $r=0.1$.}
\end{figure*}

In this section, we use the formalism of the preceding section to
determine the disentanglement. As an example, let us consider an
important class of mixed states with a single parameter $a$ like the
following \cite{Cao,Yu5,Qasimi}
\begin{equation}
\hat{\rho}_{YE}=\frac{1}{3}\left(\begin{array}{cccc}
a&0&0&0\\
0&1&1&0\\
0&1&1&0\\
0&0&0&1-a
\end{array}\right).
 \end{equation}
Apparently, the concurrence of $\rho_{YE}$ is
$\mathcal{C}_{\rho}(t)=\max\{0, K(t)\}$, and
$K(t)=|\rho_{23}(t)|-\sqrt{\rho_{11}(t)\rho_{44}(t)}$. Initially,
$\mathcal{C}(\rho(0))=\frac{2}{3}[1-\sqrt{a(1-a)}]$. In our
simulations, $\omega_0=1$ are chosen as the norm unit, and we regard
the temperature as a key factor in disentanglement process, for high
temperature $k_BT=300\omega_0$, intermediate temperature
$k_BT=3\omega_0$, and low temperature $k_BT=0.03\omega_0$,
respectively. Another reservoir parameter playing a key role in the
dynamics of the system is the ratio $r=\omega_c/\omega_0$ between
the reservoir cutoff frequency $\omega_c$ and the system oscillator
frequency $\omega_0$. As we will see in this section, by varying
these two parameters $k_BT$ and $r=\omega_c/\omega_0$, the time
evolution of the open system varies prominently from Markovian to
non-Markovian.

 In Fig.2, the time evolutions of the
non-Markovian concurrence for various values of the parameter $a$ in
low temperature is plotted. From Fig.2 we can see that the
entanglement dynamic relies on the different values of
$r=\omega_c/\omega_0$. If the spectrum of the reservoir does not
completely overlap with the frequency of the system oscillator,
$r\ll1$, we can see from Fig.2 that the ESD time is considerably
long. As increases the ratio $r$, the ESD time becomes shorter and
shorter. With different initial state we can see that the
concurrence varies prominently. When the initial state $a=0$, the
non-Markovian entanglement decay slowly, as increasing $a$, the
entanglement decay intensely, which means that we can prepare
certain initial entanglement states and use this fact to control the
system environment in order to prolong the entanglement time.

Fig.3 is the medium temperature case. Like Fig.2, under different
systems, different entanglement initial states, corresponding to
different values of $a$, and different $r$, some decay faster, some
slower. But there are some fundamental difference between Fig.2 and
Fig.3. In Section III, we get the concurrence in the long time
limit, and we affirmed that when temperature $k_BT=0$, the dynamics
of
 non-markovian
 entanglement concurrence $\mathcal{C}$ is fundamentally different from
 the case of
 $k_BT>0$. As we can see from Fig.3, for ``$\rho_{YE}$" states, as
soon as the temperature larger than zero, the concurrence always
disappear at finite time and there were no long-lived entanglement
for any value of $a$, which means that ESD must happen. The
theoretical proof is $K(t\rightarrow\infty)<0$. But when $k_BT=0$,
the stationary value of $K(t\rightarrow\infty)$ equals zero. So,
whether or not and when the ESD will happen are not sure in
$k_BT=0$. In Fig.4 we give a numerical analysis of entanglement
dynamic with different initial states and find that there exists a
$\xi\in(0,1)$, for almost all values $a>\xi$, the concurrence is
completely vanished at a finite time, which is the effect of ESD.
However, for $0\leq a\leq\xi$, the entanglement of this state decays
exponentially. But when $t\rightarrow\infty$, for all initial state,
i.e. $a\in[0, 1]$ the concurrence will tend to be $0$.

Fig.5 is the high temperature case. One of the remarkable phenomenon
in this figure is that the ESD time is short. In typical
experimental conditions, quantum dots are subjected to an external
magnetic field $B\sim1-10T$ \cite{Hanson}, the ESD time
$t_{ESD}\sim(3\times10^{-1}-3)/k_BT$. Another obvious phenomenon is
in high temperature the Markovian quantum system decays
exponentially and vanish only asymptotically, but in the
non-Markovian system the concurrence $C_{\rho}(t)$ oscillates, which
is evidently different from the Markovian. In this case the
non-Markovian property becomes evidently. This oscillatory
phenomenon is induced by the memory effects, which allows the two
qubit entanglement to reappear after a dark period of time. This
phenomenon of revival of entanglement after finite periods of
``entanglement death" appears to be linked to the environment single
qubit non-Markovian dynamics, in particular, the $\Delta(t)<0$ at
some times in some environment \cite{Maniscalco2}. The physical
conditions examined here are, moreover, more similar to those
typically considered in quantum computation, where qubits are taken
to be independent and where qubits interact with non-Markovian
environments typical of solid state microdevices \cite{Vega}.

As we indicated above, temperature is one of the key factor in the
entanglement dynamic. Figs. 2, 3, 4, 5 are plotted in the chosen
temperature,
 while in Fig. 6 $k_BT$ ranges from $0$ to $100$. In Fig. 6 the concurrence vs
``temperature $k_BT$" vs $\omega_0t$ in $r=0.1$, and the initial
state is the ``$X_{YE}$" state with $a=0$. From Fig. 6 we can
compare the non-Markovian entanglement dynamics with the Markovian
one clearly.  The left is the non-Markovian one from which we can
see the oscillation of the concurrence. Moreover, at the $0$
temperature the non-Markovian effect is faint, as the temperature
rises, the non-Markovian becomes more and more obvious, while the
Markovian one decays exponentially. This phenomenon embodies the
non-Markovian effect, which is evidently different from the
Markovian property. Maniscalco S \emph{et. al} studied the
separability function $S(\tau)$ in \cite{SM}, where entanglement
oscillation appears for twin-beam state in non-Markovian channels in
high temperature reservoirs. Both of them have the same phenomenon.
Ref. \cite{Maniscalco2} gave a distribution curve  when
$\Delta(r,t)-\gamma(r,t)>0$ and $\Delta(r,t)-\gamma(r,t)<0$. We
convince that due to the non-Markovian memory effect, particularly
$\Delta(t)<0$ in Eqs (20), the entanglement concurrence oscillates.
With $\Delta(t)-\gamma(t)>0$ the concurrence descended whilst
$\Delta(t)-\gamma(t)<0$ the concurrence ascended, which guide us to
adjust the temperature to control the entanglement evolution. In
order to show this and motivate the related research we design the
open loop controller
\begin{equation}
k_BT=e^{-\alpha|\Delta(t)-\gamma(t)|}k_BT_0
\end{equation}
where $\alpha$ is the modulation, and $k_BT_0$ is the initial
temperature. In Fig. 7, we plot the controlled entanglement
evolution, where the initial temperature is chose as $k_BT_0=30$,
which oscillates and ESD occurs at $t\approx19$. According to Fig.
1, $\gamma(t)$ can be neglected. For different modulation $\alpha$,
different controlled entanglement evolution is plotted, and the ESD
time can be prolonged for considerable long time.

\begin{figure*}
\centerline{\scalebox{0.8}[1]{\includegraphics{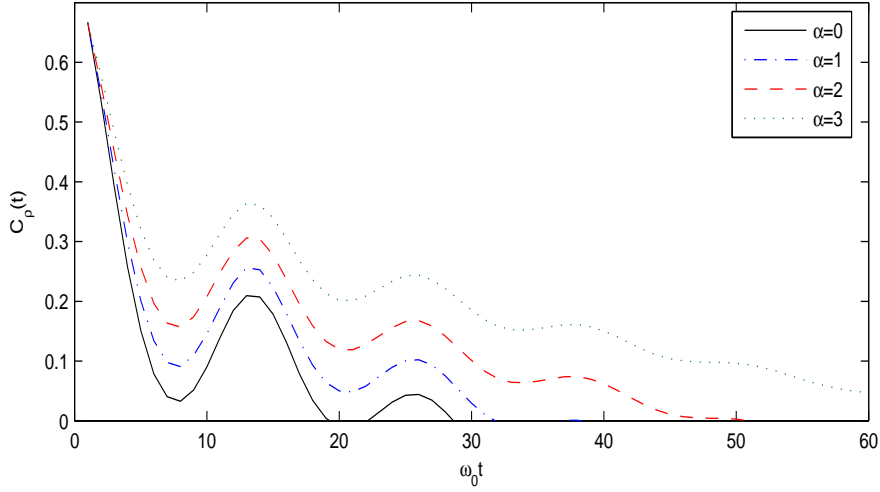}}}
\caption{(Color online)Controlled entanglement evolution with
different modulation $\alpha=3$ (green dotted line), $\alpha=2$ (red
dashed line),  $\alpha=1$ (blue dashed-dotted line), and initially
evolution (black solid line), respectively.}
\end{figure*}

\section{Conclusions}
In this paper we have presented a procedure that allows to obtain
the dynamic of a system consisting of two identical independent
qubits, each of them locally interacting with a bosonic reservoir. A
non-Markovian master equation between two qubit systems in the same
environment was obtained.  We characterize our entanglement by the
temperature and the ratio $r$ between $\omega_0$ the characteristic
frequency of the quantum system of interest, and $\omega_c$ the
cut-off frequency of Ohmic reservoir. For a broad class of initially
entangled states, ``X" states, by useing Wootters' concurrence, we
analyze the long time limit phenomenon of the entanglement dynamic.
 We find that the dynamic of
non-markovian entanglement concurrence $\mathcal{C}_{\rho}(t)$ at
temperature $k_BT=0$ is fundamentally different from $k_BT>0$.  When
$k_BT=0$, from our numerical analysis, we find that ``entanglement
sudden death" (ESD) occurs  depending on the initial state, but if
$k_BT>0$ the concurrence always disappear at finite time, which
means that ESD must happen. In the $k_BT=0$ case, we find that there
exist a $\xi\in(0,1)$, for all values $a>\xi$, the concurrence is
completely vanished in a finite time, which is the effect of ESD.
However, for $0\leq a\leq\xi$, the entanglement of this state decays
exponentially. But when $t\rightarrow\infty$, for all initial state,
i.e. $a\in[0, 1]$ the concurrence will tend to be $0$. From our
numerical analysis we also find that the entanglement dynamic relies
on the different values of $r=\omega_c/\omega_0$. If $r\ll1$, the
ESD  time is considerably long. As increases the ratio $r$, the ESD
time becomes shorter and shorter. Moreover, when the initial state
$a=0$, the non-Markovian entanglement decays slowly, as increases
$a$, the entanglement decays intensely. Most of all, we have shown
that the non-Markovian dynamics of entanglement, described by
concurrence, presents oscillation even revivals after entanglement
disappearance, typically for high temperature non-Markovian system.
At last, we design the open loop controller which adjust the
temperature to control the entanglement and prolong the ESD time.

\section*{Acknowledgments}
This work was supported by the National Natural Science Foundation
of China (No. 60774099, No. 60221301), the Chinese Academy of
Sciences (KJCX3-SYW-S01), and by the CAS Special Grant for
Postgraduate Research, Innovation and Practice.

\end{document}